\documentclass[prd,onecolumn,showpacs,floatfix,superscriptaddress,nofootinbib]{revtex4-2}
\usepackage{graphicx}
\usepackage{epsfig}
\usepackage{bm}
\usepackage{amssymb}
\usepackage{float}
\usepackage{amsmath}
\usepackage{dcolumn}
\usepackage{cancel}

\usepackage{mathrsfs}
\usepackage{dcolumn}
\usepackage{graphicx}
\usepackage{amsmath}
\usepackage{amsfonts}
\usepackage{amssymb}
\usepackage{microtype}
\usepackage{subfigure}
\usepackage{makeidx}
\usepackage{bm}
\usepackage{epsf}
\usepackage{color}
\usepackage{multirow,dcolumn}
\usepackage{graphicx}
\usepackage{mathrsfs}
\graphicspath{{Images/}}

\def\doi{http://doi.org}

\def\be{\begin{equation*}}
\def\ee{\end{equation*}}

\def\Ref{\ref}

\begin{document}

\title{Exact Black Hole Solutions with a Conformally Coupled Scalar Field and Dynamic Ricci Curvature in $f(R)$ Gravity Theories}

\author{Thanasis Karakasis}
\email{asis96kar@gmail.com} \affiliation{Physics Division,
National Technical University of Athens, 15780 Zografou Campus,
Athens, Greece.}

\author{Eleftherios Papantonopoulos}
\email{lpapa@central.ntua.gr} \affiliation{Physics Division,
National Technical University of Athens, 15780 Zografou Campus,
Athens, Greece.}

\author{Zi-Yu Tang}
\email{tangziyu@ucas.ac.cn}
\affiliation{School of Fundamental Physics and Mathematical Sciences, Hangzhou Institute for Advanced Study, UCAS, Hangzhou 310024, China}
\affiliation{School of Physical Sciences, University of Chinese Academy of Sciences, Beijing 100049, China}

\author{Bin Wang}
\email{wang\_b@sjtu.edu.cn}
\affiliation{School of Aeronautics and Astronautics, Shanghai Jiao Tong University, Shanghai 200240, China}
\affiliation{Center for Gravitation and Cosmology, College of Physical Science
and Technology, Yangzhou University, Yangzhou 225009, China}

\vspace{17.5cm}
\begin{abstract}
We report exact black hole solutions in asymptotically flat or (A)dS four-dimensional spacetime with a conformally coupled self-interacting scalar field in $f(R)$ gravity. We first consider the asymptotically flat model $f(R) = R -2\alpha \sqrt{R}$ and derive an exact black hole solution. Then, we consider the asymptotically (A)dS model $f(R) =R -2 \Lambda -2 \alpha  \sqrt{R-4 \Lambda }$ and derive an exact black hole solution. In both cases the modified gravity parameter $\alpha$, which has the dimension of the inverse mass, cannot be set to zero and the self-interacting potential is determined from the Klein-Gordon equation, preserving the conformal invariance. The thermodynamics of the solutions is also studied.
\end{abstract}
\vspace{3.5cm}

\maketitle

\flushbottom

\tableofcontents

\section{Introduction}

The first exact black hole solution with a scalar field as a matter field was found by Bocharova, Bronnikov and  Melnikov and independently by Bekenstein, called BBMB black hole \cite{BBMB}. The scalar field is conformally coupled to gravity, resulting in the vanishing of the trace of the energy-momentum tensor, which means that the scalar curvature is constant and in this particular case in the absence of cosmological constant is zero. The resulting spacetime is the external Reissner-Nordstr\"om (RN) spacetime and the scalar field diverges at the black hole horizon. It was also shown in \cite{bronnikov} that this solution is unstable under scalar perturbations. Later, a scale was introduced to the theory via a cosmological constant in \cite{Martinez:2002ru} and also a quartic scalar potential that does not break the conformal invariance of the action, which gives a very simple relation between the scalar curvature and the cosmological constant. In this case, the scalar field does not diverge at the horizon, but the solution is found to be unstable \cite{Harper:2003wt}.

Regarding the minimal coupling case, the first exact black hole solution was presented in \cite{Martinez:2004nb}, the MTZ black hole. The scalar potential is fixed ad hoc, the geometry of the solution is hyperbolic and the scalar field remains finite at the black hole horizons. In \cite{Martinez:2006an} the electrically charged case was discussed. In \cite{Kolyvaris:2009pc}, a potential that breaks the conformal invariance of the action of the MTZ black hole in the Jordan frame was considered and new black hole solutions where derived. In \cite{Gonzalez:2013aca} the scalar field was fixed ad hoc and novel black hole solutions were investigated, letting the scalar potential to be determined from the equations and in \cite{Gonzalez:2014tga} the electrically charged case is considered. In \cite{Anabalon:2012tu}, asymptotically (anti) de Sitter black holes and wormholes with a self-interacting scalar field in four dimensions were investigated. In \cite{Kolyvaris:2013zfa, Kolyvaris:2011fk, Rinaldi:2012vy, Minamitsuji:2013ura, Anabalon:2013oea} black holes with non-minimal derivative coupling were studied.  However, the scalar field which was coupled to Einstein tensor should be considered as a particle living outside the horizon of the black hole because it blows up on the event horizon. In \cite{Boudet:2020eyr} super-entropic black holes with Immirzi hair were investigated and Plebanski-Demianski solutions in quadratic gravity with conformally coupled scalar fields were also found \cite{Cisterna:2021xxq}.  Black holes in Lanczos-Lovelock gravity theories with a non-minimally coupled scalar field were recently discussed \cite{Bravo-Gaete:2021hza} and hairy black holes in disformal gravity theories were investigated in \cite{Erices:2021uyu}.

In the context of $f(R)$ gravity  (for a review see \cite{Sotiriou:2008rp, DeFelice:2010aj}) several black hole solutions have been found  recently. Vacuum solutions were discussed in \cite{Sebastiani:2010kv, Multamaki:2006zb, Hendi:2014wsa, Hendi:2014mba, Nashed:2020mnp}, while charged black hole solutions were also found in \cite{Tang:2019qiy, Elizalde:2020icc, Nashed:2020tbp, Nashed:2019uyi, Nashed:2019tuk, Cembranos:2011sr, delaCruzDombriz:2009et, Hendi:2011eg}. In \cite{Nojiri:2020blr}, dynamical black holes, characterized by time-varying apparent horizons, in various theories of gravity, including the $f(R)$ theory were investigated. The  possibility of scalarization of black holes in the context of $f(R)$ gravity was discussed in \cite{Tang:2020sjs} and $(2+1)$-dimensional black holes with a minimally coupled self-interacting scalar field in the context of $f(R)$ gravity was discussed in \cite{Karakasis:2021lnq}. Recently non-trivial black hole solutions were investigated \cite{Nashed:2021sey}. These solutions are considered non-trivial in the sense that $g_{tt} g_{rr} \neq -1$ and the integration of the field equations is achieved by fixing the $f_R(r)$ function. Finally in \cite{Nashed:2021ffk} the uniqueness of non-trivial spherically symmetric black hole solution in special classes of $f(R)$ gravitational theory was discussed.

 The viability of any theory of gravity should also be tested against the astrophysical observations. Differences between general relativity (GR) and alternative theories are expected to occur for strong gravitational fields, such as the ones created by different compact objects like neutron stars, strange stars and black holes.   In \cite{Staykov:2015cfa} an $f(R)$ modified theory of  the form $ f(R) = R + \alpha  R^{2} $ was investigated. Studying neutron stars in which strong gravity effects are non-negligible, it was found that
the neutron stars within these theories can differ significantly from their GR counterpart which makes them a very good candidate to test $f(R)$ theories on astrophysical scales \cite{Doneva:2015hsa}.

In this $f(R)$ modified theory studying   non-perturbatively and self-consistently the structure of neutron stars, bounds on the  $\alpha $ parameter were imposed \cite{Yazadjiev:2014cza}. Employing various equations of state for the neutron star, the mass-radius relations  were obtained constraining the $\alpha $ parameter resulting at large   deviations from GR. Then there was a discussion of the choice of equation of state \cite{Ozel:2010fw,Arapoglu:2010rz,Staykov:2014mwa} trying to impose a realistic equation of state  to search for predictions or derive relations that are independent of the equation of state.

In our previous work \cite{Karakasis:2021lnq}  we considered $(2+1)$-dimensional $f(R)$ gravity with a self interacting scalar field as a matter field.  The motivation for considering an explicit scalar field was  that  in $f(R)$ gravity theories  if a conformal transformation is applied from the   Jordan frame to the Einstein frame then,  a new scalar field  with a potential appears in the Lagrangian. The generated
 scalar-tensor theory has  a geometric (gravitational) scalar field which however cannot dress a $f(R)$ black hole with  hair \cite{Sultana:2018fkw, Canate, Canate:2015dda}.  Without specifying the form of the $f(R)$ function we derived the field equations and we showed  that the $f(R)$ model has a direct contribution from the scalar field.

At first we considered the case, where $f_R(r) = 1 - \int \int \phi '(r)^2 drdr$. Integrating  with respect to the Ricci scalar we  obtained a pure Einstein-Hilbert term and another term that depended  on the scalar field. The scalar curvature was dynamical and due to its complexity it was difficult to obtain an exact form of the $f(R)$ function. Using asymptotic approximations, we showed that the scalar charges make the theory to deviate form Einstein's Gravity. Then an exact black hole solution dressed with a scalar hair was found, in which the scalar charge appears in the $f(R)$ function and  the thermodynamics of the solution was studied. We further considered a pure $f(R)$ theory supported by the scalar field. We showed that the thermodynamic and observational constraints require that the pure $f(R)$ theory to be builded with a phantom scalar field. The scalar charge  is the one that determines the behaviours of the solution.

In this work we will extent our previous work to a more realistic model. We will work in $(3+1)$-dimensions with the aim to
generate a compact object with regular matter, i.e. a black hole dressed with scalar hair. The common practice in constructing such objects is to  choose a particular equation of state, expressing the matter content of the $f(R)$ theory. Instead of that, we will introduce matter as a conformably coupled scalar field and a dynamic scalar curvature.  This particular type of theory, $f(R)$ gravity and non-minimally coupled scalar fields as matter in $(3+1)$-dimensions,  has been previously considered for cosmological purposes \cite{Pi:2017gih, delaCruz-Dombriz:2016bjj}.  In our study, the scalar field is introduced in the action as a matter field in the way it was done in the GR context \cite{BBMB, Martinez:2002ru} and we study its effect on a metric ansatz solving the field equations. In \cite{Sultana:2018fkw} there is reported a black hole solution with a conformally coupled scalar field for $f(R) =R^2$, but since $R^2$ describes an exact (A)dS space \cite{Kehagias:2015ata}, the solution turns out to be identical to the one reported in \cite{Martinez:2002ru}.

 At first we will discuss the well known GR cases of the BBMB black hole \cite{BBMB} and the de Sitter black hole with a conformally coupled scalar field \cite{Martinez:2002ru}, in order to compare our modified gravity solutions with the Einstein Gravity ones. In the BBMB black hole model, the action, except for
the Ricci scalar $R$ includes the kinetic energy of the scalar field and its conformal coupling to $R$. In our model a general $f(R)$ function is introduced instead of the Ricci scalar.

In the literature there are various choices for the $f(R)$ function (for an extensive review of the $f(R)$ gravity theories see \cite{Nojiri:2010wj,Capozziello:2011et}.) These choices depend on the requirements of having consistent gravity theories, free of ghosts and tachyonic instabilities. The parameters introduced in the $f(R)$ functions are constrained by the cosmological observations. In our study we discuss the profiles $f(R) = R-2 \alpha  \sqrt{R}$ and $ f(R) =R -2 \Lambda -2 \alpha  \sqrt{R-4 \Lambda }$. These particular profiles allow us to find exact black hole solutions dressed with a scalar field. The solutions respect the conformal invariance and the self-interacting potential of the scalar field is determined from the field equations, respecting the conformal invariance. We checked the trace of the resultant energy-momentum tensor and find it vanishing, as a result of the conformal invariance. The scalar field diverges between the inner and outer horizon which is what happens in the context of GR \cite{Martinez:2002ru}. We also studied the thermodynamics of the solutions and we found that an electric charge has to be introduced, for the solution to have positive entropy.

As it happens in the astrophysical applications of the $f(R)$ gravity theories, since we are in a strong gravity regime, the  parameter $\alpha $ having dimensions of inverse mass, plays a decisive role in the construction of compact objects like black holes. Solving the full system of field equations we find that this parameter contributes to the mass of the black hole, specifies  the strength of the scalar field and influences the thermodynamical properties of the solutions.

If we switch off the scalar field, these profiles of the $f(R)$ function  can generate analytic  spherically symmetric black hole solutions that
have been extensively studied \cite{Sebastiani:2010kv,Multamaki:2006zb,Amirabi:2015aya}, while in \cite{Nashed:2019tuk} an electromagnetic field was added. It is interesting to note that these black holes are characterized by the
parameter $\alpha $ that makes solutions deviate from the standard solutions of GR. The Kretschmann
scalar and squared Ricci tensor are shown to depend on the parameter $\alpha $ which is not allowed to be zero.
Also these black hole solutions have interesting  thermodynamical properties \cite{Nashed:2019tuk}.  

This paper is organized as follows: In section \ref{sect2} we briefly discuss the black hole solutions with a conformally coupled scalar field in the context of GR. In section \ref{sect3} we derive novel black hole solutions with a conformally coupled scalar field and dynamic Ricci curvature. The thermodynamics of the solutions is also studied. Finally in section \ref{sect4} we conclude.

\section{Black Hole Solutions in Conformal Gravity Theories }
\label{sect2}

In this Section we discuss the GR black hole solutions with a scalar field conformally coupled to gravity. The first solution was found by Bocharova, Bronnikov and  Melnikov and independently by Bekenstein, called the BBMB black hole \cite{BBMB}.

Consider the action in which a scalar field is conformally coupled to gravity
\begin{equation} S = \int d^{4}x\sqrt{-g}\left[\cfrac{1}{16\pi G}R - \cfrac{1}{2}\nabla^{\mu}\phi \nabla_{\mu}\phi - \cfrac{1}{12}R\phi^{2}\right]~.  \label{BBMB_action}\end{equation}
 By variation of the above action we obtain the Einstein equation and Klein-Gordon equation respectively
\begin{eqnarray}
G_{\mu \nu} &=& 8\pi G T_{\mu \nu}~,\\
\Box \phi &=& \cfrac{1}{6}R\phi~,
\end{eqnarray}
where the energy-momentum tensor is,
\begin{equation}  T_{\mu \nu} = \nabla_{\mu}\phi\nabla_{\nu}\phi - \cfrac{1}{2}g_{\mu\nu}\nabla^{\alpha}\phi \nabla_{\alpha}\phi + \cfrac{1}{6}\big(g_{\mu\nu}\Box - \nabla_{\mu}\nabla_{\nu}  + G_{\mu\nu} \big)\phi^{2}. \end{equation}
In the following  we will set $8\pi G=1$. The matter part of the action is invariant under conformal transformations
\begin{equation} g_{\mu\nu} \to \Omega ^2 (x) g_{\mu\nu}~, \hspace{1.0cm} \phi \rightarrow \Omega^{-1}(x) \phi~. \label{conformal}\end{equation}
As a result, the energy-momentum tensor is traceless and from the Einstein equations we get $ R=0 $ for the scalar curvature.

Considering the spherically symmetric metric ansatz
\begin{equation} ds^2 = -b(r)dt^2 + b^{-1}(r)dr^2 + r^2 d\Omega^2~, \label{ds}\end{equation}
the $ R=0 $ relation gives
\begin{equation} r^2 b''(r)+4 r b'(r)+2 b(r)-2 =0~, \end{equation}
from which we get
\begin{equation} b(r) = 1 + \cfrac{c_1}{r} + \cfrac{c_2}{r^2}~, \end{equation}
where $c_1, c_2$ are constants of integration. Now, the Klein-Gordon equation becomes
\begin{equation} \Box \phi = 0~. \end{equation}
From the Einstein equations we can obtain a relation for the scalar field
\begin{equation} 2\phi'(r)^2 - \phi''(r)\phi(r)=0~, \label{scalar_equation}\end{equation}
which gives
\begin{equation}  \phi(r) = \cfrac{1}{c_3 r + c_4}~, \label{scalar_profile1} \end{equation}
where $c_3, c_4$ are constants of integration. Substituting the expressions for the metric function and the scalar field to Einstein and Klein-Gordon equations we obtain the  BBMB black hole
\begin{eqnarray}
b(r) &=& \left(1-\cfrac{m}{r}\right)^2 = 1-\cfrac{2m}{r} +\cfrac{m^2}{r^2}~, \label{BBMB_metric}\\
\phi (r) &=& \pm \sqrt{6}\cfrac{m}{r-m}~, \label{scal11}
\end{eqnarray}
where, $m$ is the mass of the black hole.
To summarize, the metric  function has an extra  $m^2/r^2 $ term because of the presence of the scalar field which resembles the extermal RN case, where the scalar field diverges at the black hole horizon while the Kretschmann scalar is  divergent at the origin indicating a physical singularity. Also note that in some sense the scalar field provides the mass of the black hole.

The thermodynamical properties of the BBMB black hole are interesting. The Hawking temperature  is zero, since it is given by
\begin{equation} T(r_h) = \cfrac{b'(r_h)}{4\pi} = \frac{m}{2 \pi  r^2} \left(1-\frac{m}{r}\right)\Big|_{r=m}=0~, \end{equation}
while the  entropy \cite{Barlow:2005yd, Winstanley:2004ay} becomes infinite
\begin{equation} S = \pi r_h^2\left(1 - \cfrac{1}{6}\phi(r_h)^2\right)~, \end{equation}
since the scalar field diverges at the event horizon. Therefore, the BBMB black hole does not have the  conventional thermodynamic properties of black holes as it was discussed in \cite{Zaslavskii:2002zv} and since black holes should have non-zero surface gravity (temperature) one may argue that the BBMB black hole is not a black hole.

Hence, the BBMB black hole solution is not a hairy black hole because the conformal symmetry does not allow the formation of finite hair at the horizon with a scalar charge independent of the black hole mass that can be detected asymptotically.

A generalization of the  BBMB black hole solution was presented in \cite{Martinez:2002ru}. Consider the action
\begin{equation} S =\int d^4 x \sqrt{-g}\Bigg(\cfrac{R-2\Lambda}{16\pi G} - \cfrac{1}{2}g^{\mu\nu}\partial_{\mu}\phi\partial_{\nu}\phi - \cfrac{1}{12}R\phi^2 - \alpha \phi^4\Bigg)~. \end{equation}
The resulting field equations are
\begin{eqnarray}
G_{\mu\nu} + \Lambda g_{\mu\nu} - 8\pi G T_{\mu\nu}&=& 0 ~, \\
\Box \phi - \cfrac{1}{6}R\phi - 4\alpha \phi^3 &=& 0~,
\end{eqnarray}
where
\begin{equation}  T_{\mu \nu} =\nabla_{\mu}\phi\nabla_{\nu}\phi - \cfrac{1}{2}g_{\mu\nu}\nabla^{\alpha}\phi \nabla_{\alpha}\phi + \cfrac{1}{6}\big(g_{\mu\nu}\Box - \nabla_{\mu}\nabla_{\nu}  + G_{\mu\nu} \big)\phi^{2} - \alpha g_{\mu\nu}\phi^4~.
\end{equation}
The matter part of the action is invariant under the conformal transformations (\Ref{conformal}). As a result, the energy-momentum tensor is traceless and in the presence of the cosmological constant the scalar curvature is
\begin{equation} R = 4\Lambda~. \label{curv} \end{equation}
With the  metric  (\Ref{ds}) the constant curvature relation (\Ref{curv}) gives the metric function
\begin{equation} b(r) = -\frac{\Lambda  r^2}{3}+\frac{c_2}{r^2}+\frac{c_1}{r}+1~.\label{metri} \end{equation}
From Einstein equations, we can obtain equation (\Ref{scalar_equation}) for the profile of the scalar field. Plugging the metric (\ref{metri}) and the scalar (\Ref{scalar_profile1}) back to Einstein equations, the integration constants take particular values and the solution becomes
\begin{eqnarray}
\phi(r) &=&\pm \sqrt{\frac{3}{\pi }}\frac{ \sqrt{G} M}{2 G M-2 r}~,\\
b(r) &=& 1 -\frac{\Lambda  r^2}{3} -\frac{2 G M}{r} + \frac{G^2 M^2}{r^2} =-\frac{\Lambda  r^2}{3} +\left( 1 -\cfrac{GM}{r}\right)^2~. \label{metric_for_GRdS}
\end{eqnarray}

In order to respect the conformal invariance, the parameter $\alpha$ is specified, so the solution exists only for  $\alpha =-2 \pi  G \Lambda/9$. There are three horizons the inner, event and cosmological horizon and all possible divergencies of the curvature invariants, the metric function and the scalar field are hidden behind the event horizon. We note that we cannot have a black hole solution for an AdS spacetime, since equation (\Ref{metric_for_GRdS}) is always positive.

 The thermodynamics of this solution have been discussed in \cite{Barlow:2005yd}. The temperature of the black hole is given by
\begin{equation} T = \cfrac{1}{2\pi l}\sqrt{1 - \cfrac{4M}{l}}~, \label{tempds} \end{equation}
where $l$ is the dS radius $\Lambda = 3/l^2$. The entropy at the black hole horizon is negative, while the entropy at the cosmological horizon is positive and the total entropy of the black hole is zero, since the two entropies have the same absolute value.

The black hole solution discussed in \cite{Martinez:2002ru} is a generalization of the BBMB black hole solution in the presence of a positive cosmological constant. This modification allows the scalar field to be finite on the event horizon dressing the black hole with secondary scalar hair \cite{Herdeiro:2015waa}, but still the thermodynamic properties of the solution indicate that the produced compact object does not have a conventional thermodynamic behavior.  In an attempt to understand better the thermodynamical properties of the solution a charge was introduced to the theory \cite{Martinez:2002ru, Barlow:2005yd}.

\section{Black Hole Solutions in Conformal $f(R)$ Gravity Theories}
\label{sect3}

As we have seen in the previous section in conformal gravity theories the Ricci scalar plays an important role to obtain exact black holes. If it is zero then no black hole with finite scalar hair can be produced, since the scalar field is divergent at the event horizon. If it is proportional to a positive cosmological constant then an exact black hole is generated with a  scalar field regular on the event horizon. However, in both cases we do not get a black hole with the conventional thermodynamic properties. In this section we will investigate the case of introducing a non-linear curvature correction term to the Ricci scalar in the context of the $f(R)$ gravity theory and study the thermodynamic properties of the resulting black hole solutions.

We consider the action
\begin{equation} S = \int d^4x \sqrt{-g} \left(\cfrac{f(R)}{2} - \cfrac{1}{2}\partial^{\mu}\phi\partial_{\mu}\phi - \cfrac{1}{12}R\phi^2 - V(\phi)\right)~, \label{action} \end{equation}
which consists of an arbitrarily differentiable function of the Ricci scalar and a self-interacting, non-minimally coupled to gravity scalar field. The factor of the non-minimal coupling is the conformal coupling factor and the potential is arbitrary. We will determine the potential from the field equations.
The field equations are
\begin{eqnarray}
f_R R_{\mu\nu} - \cfrac{1}{2}f(R)g_{\mu\nu} + g_{\mu\nu}\Box f_R - \nabla_{\mu}\nabla_{\nu}f_R &=& T_{\mu\nu}~, \label{EE} \\
 \Box \phi - \cfrac{1}{6}R\phi - V'(\phi) &=&0~, \label{KG}
\end{eqnarray}
where $f_R=\cfrac{df(R)}{dR}$. The energy-momentum tensor is given by
\begin{equation} T_{\mu\nu}^{\phi} = \partial_{\mu}\phi \partial_{\nu}\phi - \cfrac{1}{2}g_{\mu\nu}\partial^{\alpha}\phi\partial_{\alpha}\phi + \cfrac{1}{6}\Big(g_{\mu\nu}\Box - \nabla_{\mu}\nabla_{\nu}+ G_{\mu\nu}\Big)\phi^2 -g_{\mu\nu}V(\phi)~. \label{energy_momentum} \end{equation}
Considering  the same metric ansatz as in the GR cases (\Ref{ds}), the components of the Einstein field equation are
 \begin{equation}
0= r \left(r \left(b'( \left(\phi  \phi '-3 f'_R\right)+3 f_R b''-b \left(6 f''_R+\phi '^2-2 \phi  \phi ''\right)+3 f-6 V\right)+b' \left(6 f_R+\phi ^2\right)-12 b f'_R+4 b \phi  \phi '\right) (b-1) \phi ^2~,\label{tt}
\end{equation}
\begin{equation}
0= r \left(r \left(b' \left(\phi  \phi '-3 f'_R\right)+3 f_R b''+3 b \phi '^2+3 f-6 V\right)+b' \left(6 f_R+\phi ^2\right)-12 b f'_R+4 b \phi  \phi '\right)+(b-1) \phi ^2~, \label{rr} \end{equation}
\begin{equation}0= r \left(-2 \left(6 r b' f'_R+b \left(6 f'_R+6 r f''_R+r \phi '^2\right)\right)+4 \phi  \left(\left(r b'+b\right) \phi '+r b \phi ''\right)+\phi ^2 \left(2 b'+r b''\right)\right)+12 f_R \left(r b'+b-1\right) + 6 r^2 (f-2 V)~.\label{uu}
\end{equation}
The Klein-Gordon equation (\Ref{KG}) for the metric (\Ref{ds}) reads
\begin{equation} \frac{\phi (r) \left(r^2 b''(r)+4 r b'(r)+2b(r)-2\right)}{6 r^2}+b'(r) \phi '(r)+\frac{2 b(r) \phi '(r)}{r}+b(r) \phi ''(r)-\frac{V'(r)}{\phi '(r)} =0~. \label{KG1} \end{equation}
The trace of Einstein equation (\Ref{EE}) in tensor and differential form reads
\begin{equation} R f_R - 2f(R) + 3\square f_R = \phi \Box \phi - R\phi^2/6 - 4V(\phi)~, \end{equation}
\begin{multline}0= r \left(2 b' \left(-9 r f'_R+3 r \phi  \phi '+2 \phi ^2\right)+r \phi ^2 b''\right)+6 f_R \left(r^2 b''+4 r b'+2 b-2\right)+2 b \left(-9 r \left(2 f'_R+r f''_R\right)+3 r \phi  \left(2 \phi '+r \phi ''\right)+\phi ^2\right)\\+12 r^2 (f-2 V)-2 \phi^2~,  \end{multline}
while the trace of the energy-momentum tensor is
\begin{equation}T^{\mu}_{\mu} = \frac{\phi (r) \left(\phi (r) \left(r^2 b''(r)+4 r b'(r)+2 b(r)-2\right)+6 r \left(\left(r b'(r)+2 b(r)\right) \phi '(r)+r b(r) \phi ''(r)\right)\right)}{6 r^2}-4 V(r)~. \label{trace}\end{equation}

We will determine if the resulting solution is conformally coupled to gravity by examining if the trace (\Ref{trace}) vanishes.

In the following we will fix the $f(R)$ function, determine the resulting black hole solutions and study their thermodynamic properties.

\subsection{$f(R) = R-2 \alpha  \sqrt{R}$}

\subsubsection{Neutral Solution}

We first consider the $f(R)$ model
\begin{equation} f(R) = R-2 \alpha  \sqrt{R}~, \end{equation}
where a non-linear curvature correction is added to the Einstein-Hilbert term through the model parameter $\alpha$ which has the dimensions of $[L]^{-1}$ (inverse mass). Solving equations (\Ref{tt}), (\Ref{rr}), (\Ref{uu}) and (\Ref{KG1}) we obtain the following configurations
\begin{eqnarray}
b(r) &=& \frac{1}{2}-\frac{1}{3 \alpha  r}+ \frac{3}{64 \alpha ^2 r^2} \label{metric11}~,\\
f_R(r) &=& 1- \alpha r~,\\
f(r) &=& \frac{1-2 \alpha  r}{r^2}~,\\
\phi(r) &=& \frac{\sqrt{6}}{1-4 \alpha  r}~,\\
V(r) &=& -\frac{4 \alpha ^2}{(4 \alpha  r-1)^4}~,\\
V(\phi) &=& -\frac{1}{9} \alpha ^2 \phi ^4~,\\
R(r) &=& \cfrac{1}{r^2} \label{R1} ~.
\end{eqnarray}
The trace of the resulting energy-momentum tensor is zero, meaning that the scalar field is conformally coupled to gravity and the scalar potential which is obtained from the Klein-Gordon equation preserves the conformal invariance. The metric function has two roots
\begin{equation} r_{\pm} = \frac{8 \pm\sqrt{10}}{24 \alpha }~, \end{equation}
which are both positive for $\alpha >0$, $r_-$ being the inner while $r_+$ being the outer horizon. There exists a simple pole (divergence) in the scalar field function that lies between the horizons
\begin{equation} r_- <r_{divergence} < r_+~. \label{divergence}\end{equation}
The scalar field is finite at the event horizon and takes the value
\begin{equation} \phi(r_+) = -\frac{6 \sqrt{6}}{\sqrt{10}+2}~. \end{equation}
The behavior of the scalar field is similar with the GR case \cite{Martinez:2002ru}.
The model is stable only for $\alpha>0$ since then we have $f_{RR} = \cfrac{\alpha }{2 R^{3/2}}>0$ avoiding tachyonic instabilities \cite{DeFelice:2010aj} which means that for $\alpha >0$ we can interpret this solution as a black hole solution with a conformally coupled scalar field as matter in a viable $f(R)$ model. Therefore we will impose the condition $\alpha >0$ throughout the paper.

 If the parameter $\alpha$ is non-zero then the Ricci curvature $R$ receives a non-linear correction term and the field equations have the solution (\ref{metric11})-(\ref{R1}) in which the only free parameter is $\alpha$.  If $\alpha$ is zero then the $f(R)$ is  given by the Ricci curvature $R$ and then the field equations, as we discussed in the previous section, give the BBMB  solution (\ref{BBMB_metric}), (\ref{scal11}). In this solution the mass is an integration constant which is fixed by the gravitational potential and a mass-squared term appears because of the presence of the scalar field. In the case of the $f(R)$ theory in which gravity is stronger we have a similar behaviour. However,  the mass of the black hole  is proportional to $1/ \alpha$ and also this parameter gives an effective charge term  to the scalar field. A similar behaviour is found in the vacuum and charged black hole solutions \cite{Sebastiani:2010kv, Elizalde:2020icc, Nashed:2019tuk}. We note here that in modified theories of gravity  the strong-field gravity effects can potentially introduce much greater differences in the parameters of local solutions in these theories. These theories can pass the present weak-field gravitational tests and exhibit non-perturbative strong-field deviations away from GR in systems involving various compact objects \cite{Damour:1993hw, Sotani:2004rq}.

\begin{figure}[H]
\centering
	\includegraphics[width=.40\textwidth]{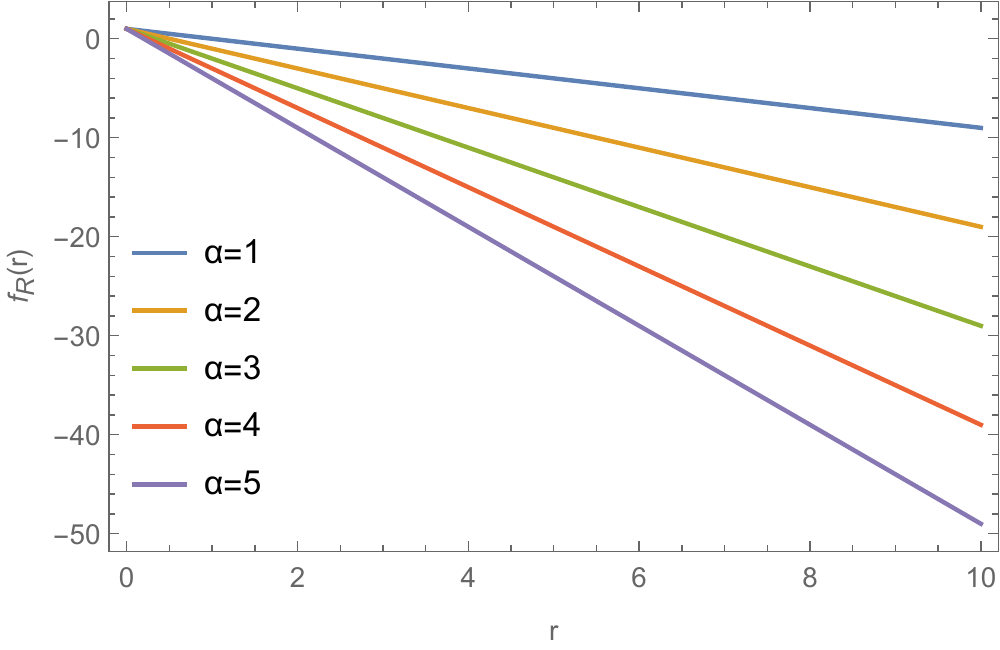}
	\includegraphics[width=.40\textwidth]{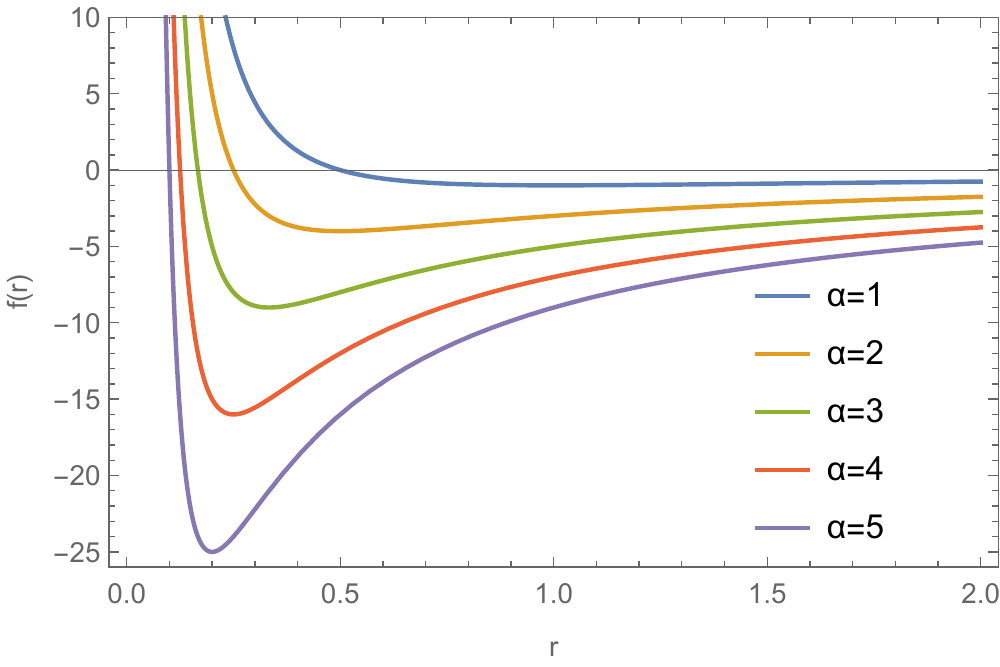}
\caption{Quantities related to the $f(R)$ model: $f_R(r)$ and $f(r)$ for different values of the parameter $\alpha$.}\label{fig.FIG1}
\end{figure}
\begin{figure}[H]
\centering
	\includegraphics[width=.40\textwidth]{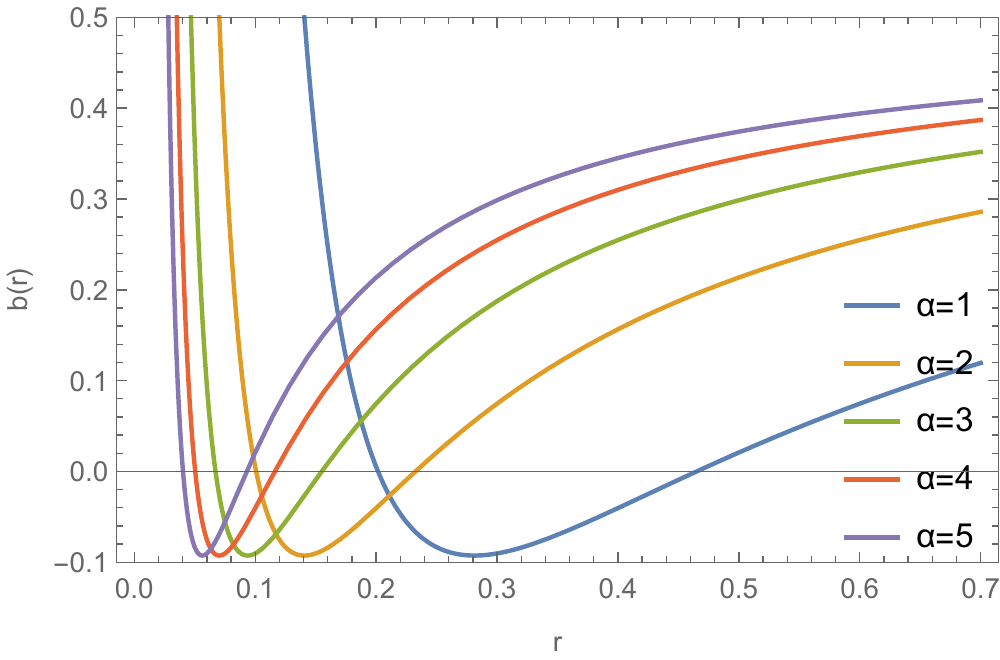}
	\includegraphics[width=.40\textwidth]{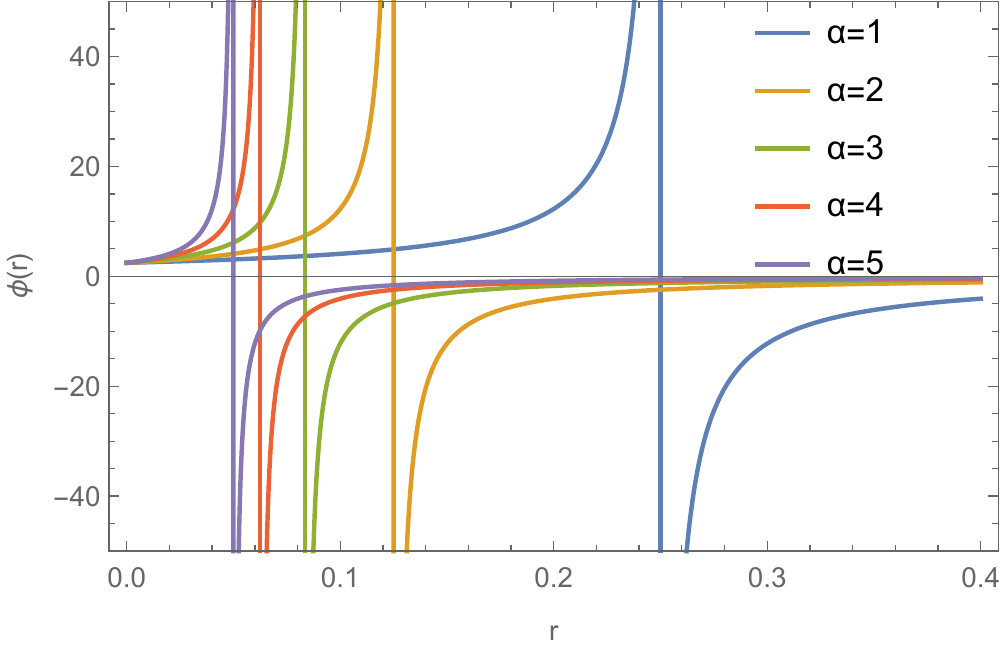}
	\includegraphics[width=.40\textwidth]{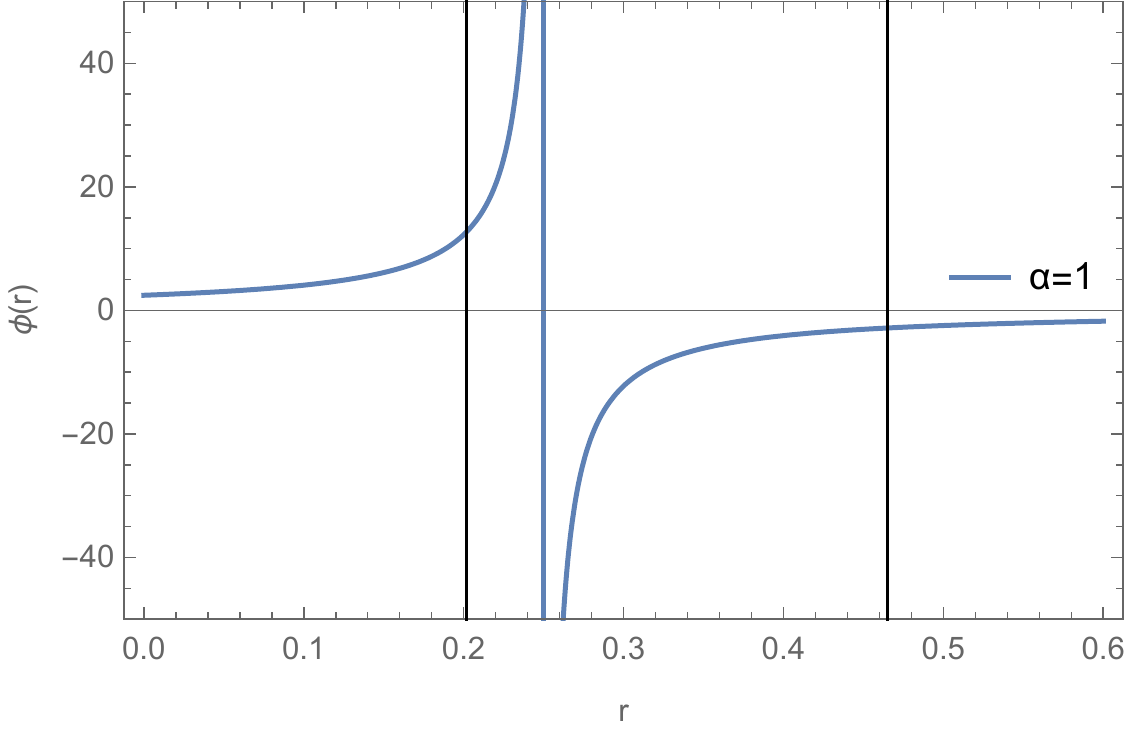}
\caption{The metric function $b(r)$ (left) and the scalar field $\phi(r)$ (right) for different values of the parameter $\alpha$. Below, we plot the scalar field while the vertical lines represent the positions of the horizons for a particular value of $\alpha$.} \label{fig.FIG2}
\end{figure}

\begin{figure}[H]
\centering
	\includegraphics[width=.40\textwidth]{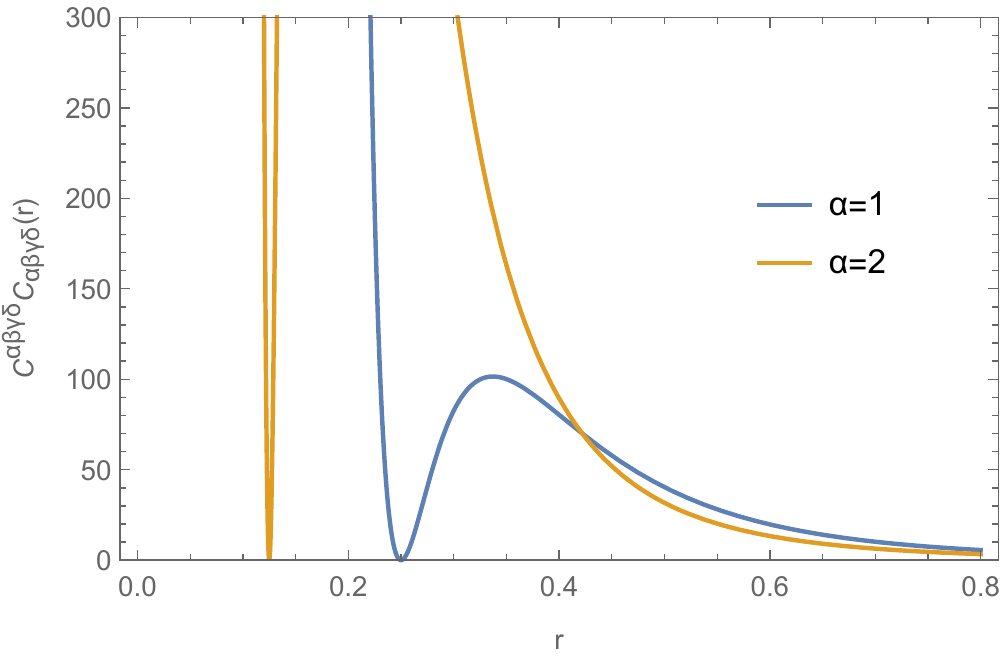}
		\includegraphics[width=.40\textwidth]{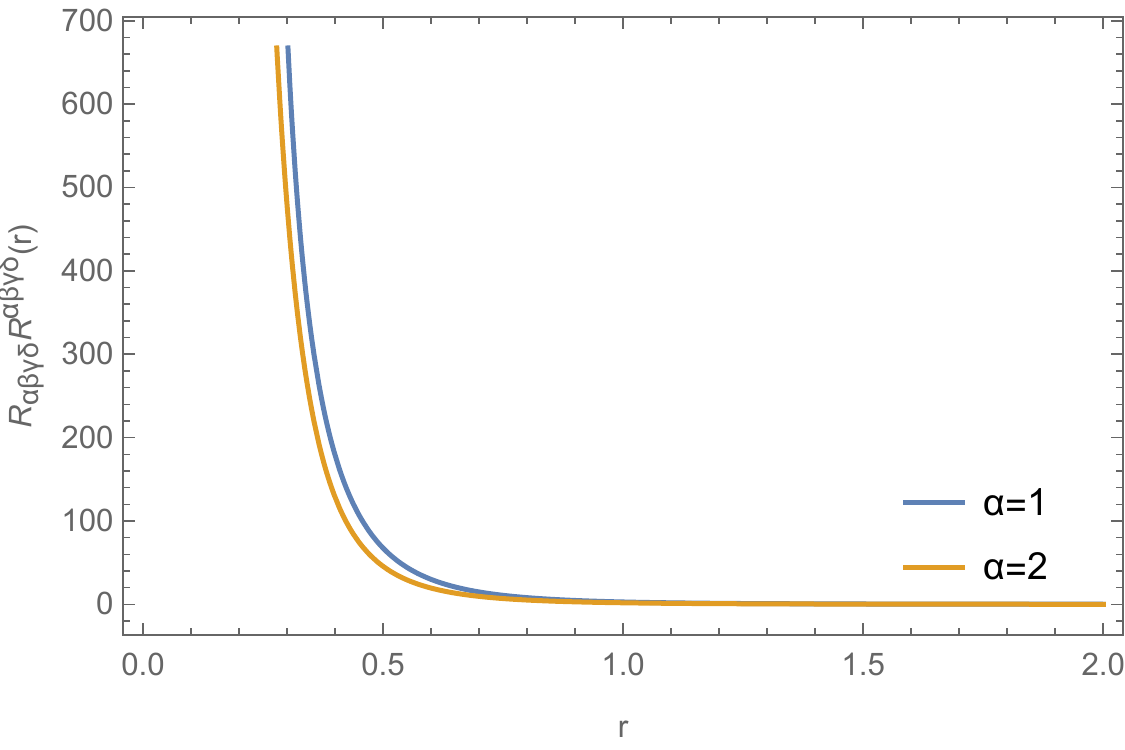}
\caption{The norm of the Weyl tensor (left) and the Kretschmann scalar (right) as functions of $r$ for different values of the parameter $\alpha$.} \label{invar}
\end{figure}
\begin{figure}[h]
\centering
\includegraphics[width=.40\textwidth]{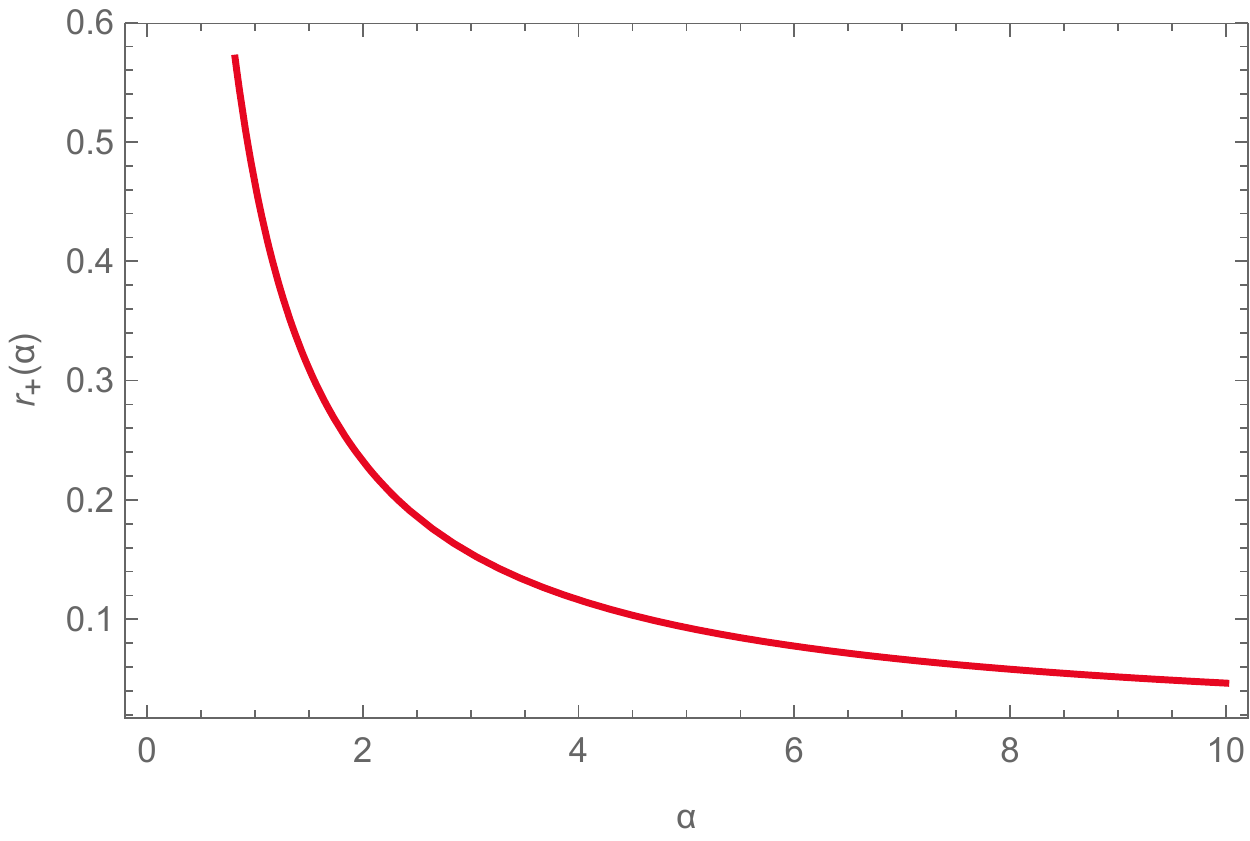}
\caption{The event (outer) horizon $r_+$ as a function of the parameter $\alpha$.} \label{r_+}
\end{figure}
In FIG. \ref{fig.FIG1} we plot the $f_{R}(r)$ and $f(r)$ (the gravitational model as a function of the radial coordinate). From $f(r)$ we can see that the gravitational effects are stronger near the origin, since the Ricci scalar diverges there, while tend rapidly to zero at large distances.
In FIG. \ref{fig.FIG2} we plot the metric function and the scalar field in order to see the behaviour of these functions in the regime (\Ref{divergence}). The metric function diverges at the origin while tends rapidly to $1/2$ at large distances. The scalar field diverges only between the inner and outer horizons and vanishes at large distances. We should note that the scalar field does not dress the black hole with some kind of hair. The only free parameter of the system is $\alpha$. The scalar field does not have a scalar charge because of the conformal invariance but, it acquires a charge from the curvature sector of the action.  We compute the Kretschmann scalar for the metric (\Ref{metric11}) and the norm of the Weyl tensor.
\begin{eqnarray}
R^{\alpha\beta\gamma\delta}R_{\alpha\beta\gamma\delta} &=& \frac{55}{48 \alpha ^2 r^6}-\frac{3}{4 \alpha ^3 r^7}+\frac{63}{512 \alpha ^4 r^8}+\frac{4}{3 \alpha  r^5}+\frac{1}{r^4}~,\\
C^{\alpha\beta\gamma\delta}C_{\alpha\beta\gamma\delta} &=& \frac{(1-4 \alpha  r)^2 (4 \alpha  r+9)^2}{768 \alpha ^4 r^8}~. \label{weyl}
\end{eqnarray}
Both scalars diverge at the origin meaning that $r \rightarrow 0$ is a physical singularity. The norm of the Weyl tensor also vanishes at the point where the scalar field diverges. We present plots for the curvature invariants in FIG. \ref{invar}. We also plot the event (outer) horizon of the black hole as a function of the modified gravity parameter $\alpha$ in FIG. \ref{r_+},  where we can see that for bigger values of $\alpha$, the black hole is formed closer to the origin of the coordinates.

\begin{figure}[H]
\centering
 \includegraphics[width=.40\textwidth]{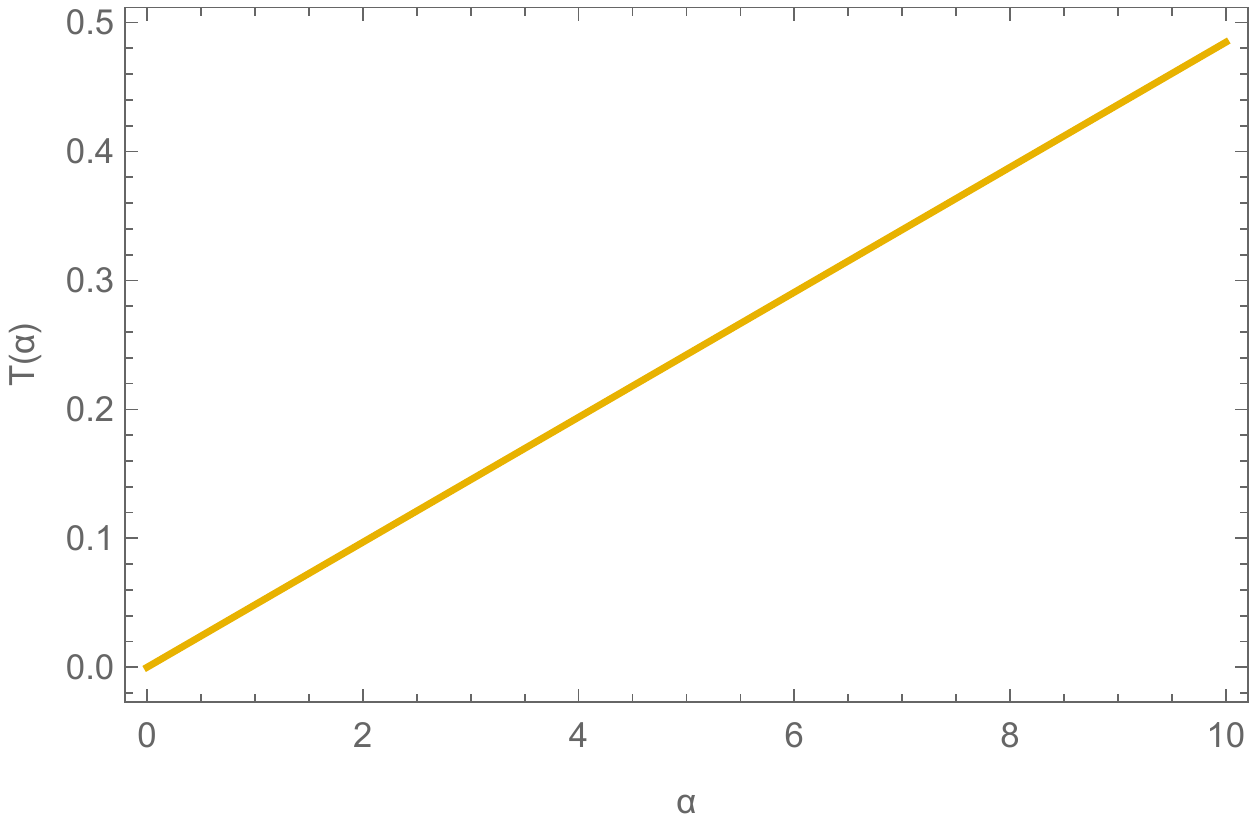}
     \includegraphics[width=.40\textwidth]{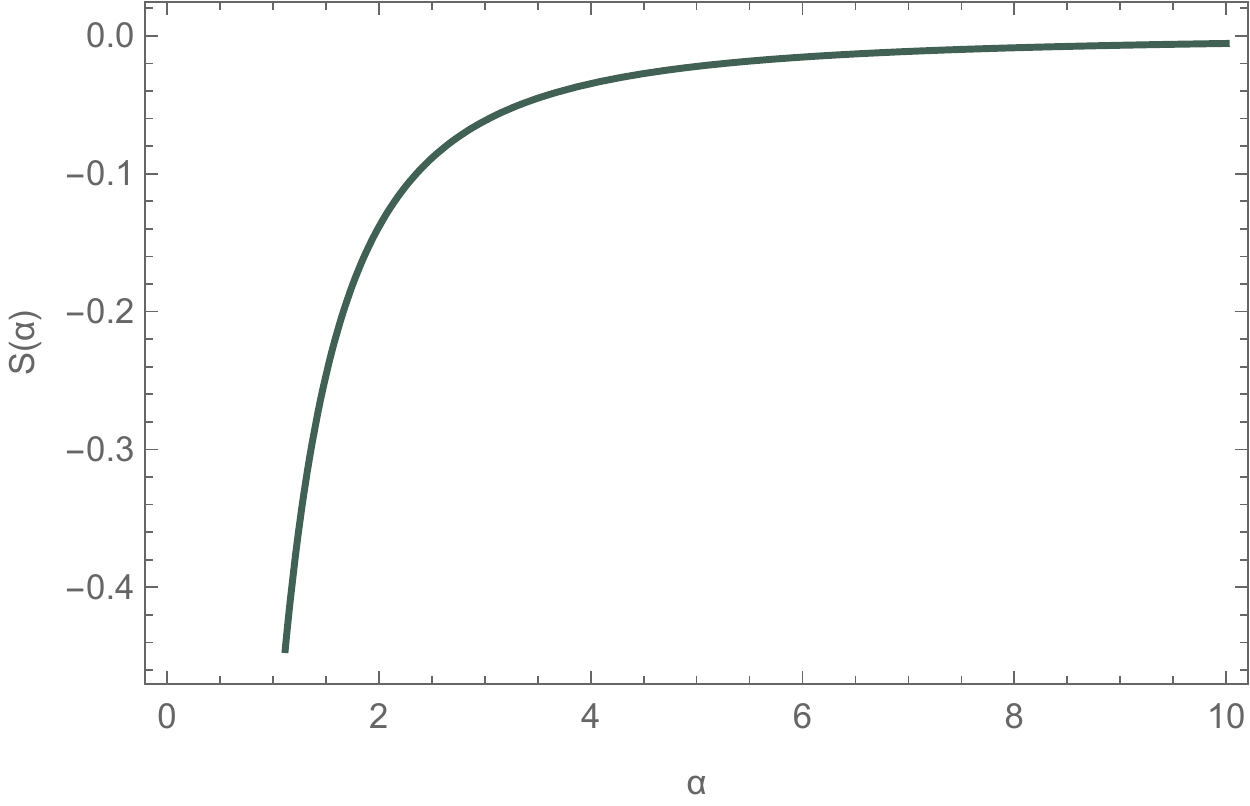}
\caption{The Hawking temperature $T(r_+)$ (left) and the Bekenstein-Hawking entropy $S(r_+)$ (right) at the event horizon of the black hole as functions of the modified gravity parameter $\alpha$.} \label{temp1_entro1}
\end{figure}

We have seen that the black hole solutions in GR with conformally coupled scalar field do not have a conventional thermodynamical behavior, due to the existence (and therefore modification of the area law for the entropy) of the non-minimally coupled scalar field. We will discuss the thermodynamics of the asymptotically flat black hole solution in the $f(R)$ gravity theory with the choice $f(R) = R-2 \alpha  \sqrt{R}$.
To compute the Hawking temperature and the Bekenstein-Hawking entropy we use the following relations \cite{Zheng:2018fyn, Nashed:2019tuk,Barlow:2005yd}
\begin{eqnarray}
T(r_+) &=&\cfrac{b'(r_+)}{4\pi}~, \label{T}\\
S(r_+) &=&\cfrac{\mathfrak{A}f_R(r_+)}{4} = \cfrac{\mathfrak{A}}{4}\Big(f_{R}^{(\text{gravity})}(r_+) + f_{R}^{(\text{matter})}(r_+)\Big) = \cfrac{\mathfrak{A}}{4}\left(1 - \alpha r_+ - \cfrac{1}{6}\phi(r_+)^2\right)~.\label{S}
\end{eqnarray}
which are the Hawking temperature and the Bekenstein-Hawking entropy, where $\mathfrak{A}$ denotes the area of the black hole $\mathfrak{A}=4\pi r_+^2$. It is of major importance to note the fact that the entropy acquires a multiplicative factor due to the existence of the non-minimal coupling between Ricci scalar and scalar field. So, we do not only have the modified gravity part but we also have contribution from the matter part of the action, that modifies the area law for the entropy. The concrete expressions of them are
\begin{eqnarray}
T(r_+) &=& \frac{32 \alpha  r_+-9}{384 \pi  \alpha ^2 r_+^3} = \frac{\left(37 \sqrt{10}-80\right) \alpha }{243 \pi }~, \label{T1}\\
S(r_+) &=& \pi  r_+^2 \left(1-\alpha  r_+-\frac{1}{\left(1-4 \alpha  r_+\right){}^2}\right) = \frac{5 \left(191 \sqrt{10}-848\right) \pi }{6912 \alpha ^2}~.
\end{eqnarray}
It is clear that we cannot set the parameter $\alpha$ to zero. The Hawking temperature is always positive and proportional to the model parameter $\alpha$ while the Bekenstein-Hawking entropy is always negative and inversely proportional to $\alpha^2$. Their figures are present in FIG. \ref{temp1_entro1}. The thermodynamics of this solution if we set $\alpha \to 0$ behaves similar to the BBMB black hole which we previously discussed.

\subsubsection{Solution with Charge}

We now add in (\Ref{action}) a Maxwell term and the whole action reads
\begin{equation} S =\int d^4x \sqrt{-g} \left(\cfrac{f(R)}{2} - \cfrac{1}{2}\partial^{\mu}\phi\partial_{\mu}\phi - \cfrac{1}{12}R\phi^2 - V(\phi)- \cfrac{1}{4}F^2\right)~, \label{charged_action}\end{equation}
where
\begin{eqnarray}
F^2 &=& F^{\mu\nu}F_{\mu\nu}~,\\
F_{\mu\nu} &=& \partial_{\mu}A_{\nu} - \partial_{\nu}A_{\mu}~.
\end{eqnarray}
By variation with respect to the inverse metric tensor, the scalar field and the $U(1)$ field we obtain the Einstein equation (\Ref{EE}), the Klein-Gordon equation (\Ref{KG}) and the Maxwell equation
\begin{equation} \nabla_{\mu}F^{\mu\nu} =0~, \end{equation}
while the energy-momentum tensor is now
\begin{equation} T^{total}_{\mu\nu} =  T_{\mu\nu}^{\phi} + T^{EM}_{\mu\nu}~,\end{equation}
where
\begin{equation} T^{EM}_{\mu\nu} =F_{\mu\alpha}F^{\alpha}_{\nu} - \cfrac{1}{4}g_{\mu\nu}F^2~.  \end{equation}
Imposing the same metric ansatz (\Ref{ds}) and the following ansatz for the electromagnetic field, allowing only radial electric fields
\begin{equation} A_{\mu} = (\mathcal{A}(r),0,0,0)~, \end{equation}
we get
\begin{equation}-\frac{2 \mathcal{A}'(r)}{r}-\mathcal{A}''(r) =0 \rightarrow \mathcal{A}(r) = \cfrac{Q}{r}~, \label{charge}\end{equation}
where $Q$ is the charge of the black hole.

Now, solving Einstein equations and (\Ref{KG1}) the only functions that change are
\begin{eqnarray}
\phi(r) &=& -\frac{\sqrt{6 -64 \alpha ^2 Q^2}}{ (4 \alpha  r-1)}~, \label{charged_phi}\\
V(r) &=& \frac{4 \alpha ^2 \left(32 \alpha ^2 Q^2-3 \right)}{3 (1-4 \alpha  r)^4}~,\\
V(\phi) &=& \frac{\alpha ^2 \phi ^4}{96 \alpha ^2 Q^2-9}~,
\end{eqnarray}
while all other functions remain the same. For the vanishing of $Q$ the solution turns back to the uncharged one. The electric charge does not appear in the metric function but modifies the scalar field and scalar potential, which is exactly what happens in the GR case \cite{BBMB, Martinez:2002ru}.

In order to have a real valued scalar field we should impose the condition
\begin{equation} -\frac{1}{4\alpha} \sqrt{\frac{3}{2}} <Q<\frac{1}{4\alpha} \sqrt{\frac{3}{2}}~. \end{equation}
The scalar field still remains finite at the black hole horizon and now takes the value
\begin{equation} \phi(r_+) = -\frac{6 \sqrt{6-64 \alpha ^2 Q^2}}{\sqrt{10}+2}~. \end{equation}
Since the metric function remains unchanged, the divergence of the scalar field lies in between of the inner and event horizon of the black hole.

Thermodynamically, the temperature will be still given by equation (\Ref{T1}), since the electric charge does not appear in the metric function. The entropy will change though, now the electric charge has an impact on the profile of the scalar field. The entropy at the event horizon of the black hole will be given by (\Ref{S})
\begin{equation} S(r_+) = \frac{2}{3} \left(11-2 \sqrt{10}\right) \pi  Q^2+ \frac{5 \left(191 \sqrt{10}-848\right) \pi }{6912 \alpha ^2}~, \label{charged_entropy}\end{equation}
where we have taken into account that the event horizon is located at $r_{+} =(8 +\sqrt{10})/24 \alpha$. The second term in the relation (\Ref{charged_entropy}) is the entropy of the uncharged black hole.  It is clear that the addition of the electric charge gains more entropy for the black hole throught the scalar field. The entropy is positive when the electric charge and the modified gravity parameter $\alpha$ are related throught the inequalities
\begin{equation} \frac{\sqrt{10}}{\alpha} \sqrt{\frac{191 \sqrt{10}-848}{\left(2 \sqrt{10}-11\right)  }}+96 Q<0 \hspace{0.5cm}\text{or}\hspace{0.5cm} \frac{\sqrt{10}}{\alpha} \sqrt{\frac{191 \sqrt{10}-848}{\left(2 \sqrt{10}-11\right)  }}<96 Q \hspace{0.2cm}\&\hspace{0.2cm} \frac{\sqrt{6}}{\alpha }>8 Q~.\end{equation}

\begin{figure}[h]
\centering
\includegraphics[width=.40\textwidth]{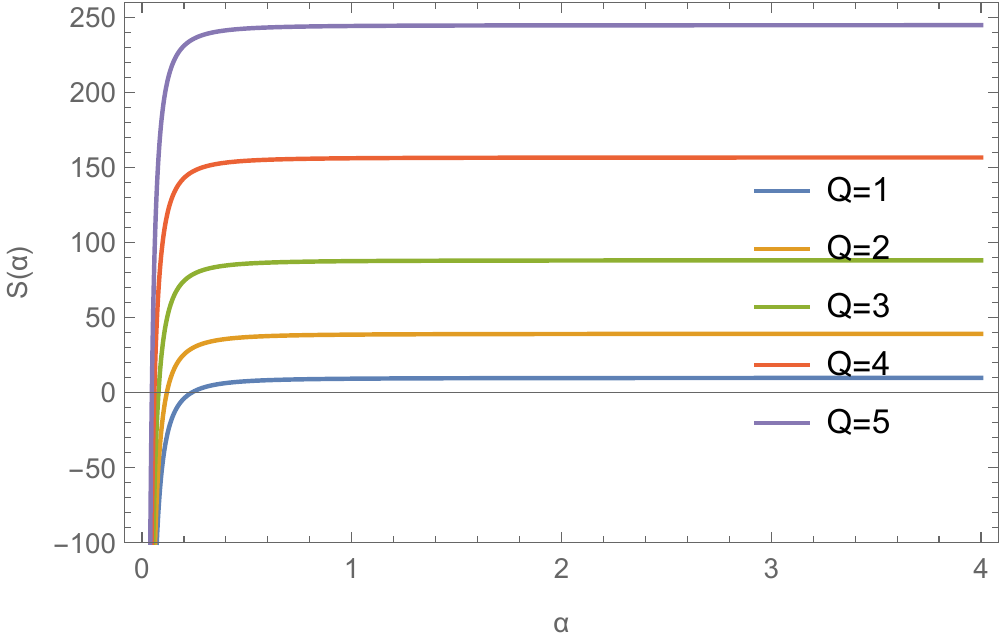}
\caption{The Bekenstein-Hawking $S(r_+)$ entropy of the charged solution at the event horizon of the black hole as a function of $\alpha$ while changing the electric charge $Q$.} \label{charged_S}
\end{figure}

In FIG. \Ref{charged_S} we give a plot of the entropy at the event horizon of the black hole as a function of the modified gravity parameter $\alpha$ while changing the electric charge. We can see that when the above relations hold, the entropy becomes positive while otherwise is negative.

\subsection{$ f(R) =R -2 \Lambda -2 \alpha  \sqrt{R-4 \Lambda }$}

\subsubsection{Neutral Solution}

In this Section we introduce a cosmological constant in the $ f(R)$ function
\begin{equation} f(R) =R -2 \Lambda -2 \alpha  \sqrt{R-4 \Lambda } \label{F2}~. \end{equation}
In this case the solution to the system of equations (\Ref{tt}), (\Ref{rr}), (\Ref{uu}) and (\Ref{KG1}) yields the following configurations
\begin{eqnarray}
f_R(r) &=&1-\alpha  r~, \label{cosmo_f_R}\\
b(r) &=&\frac{1}{2}-\frac{1}{3 \alpha  r} + \frac{3}{64 \alpha ^2 r^2}-\frac{\Lambda  r^2}{3}, \label{cosmo_b}\\
f(r) &=& \frac{2 \Lambda  r^2-2 \alpha  r+1}{r^2}~, \label{cosmo_f(r)}\\
\phi(r) &=& \frac{\sqrt{6}}{1-4 \alpha  r}~,\\
V(r) &=& -\frac{4 \alpha ^2+\Lambda }{(4 \alpha  r-1)^4}~,\\
V(\phi) &=&-\frac{1}{36} \phi ^4 \left(4 \alpha ^2+\Lambda \right)~,\\
R(r) &=& 4 \Lambda +\frac{1}{r^2}~. \label{cosmo_R(r)}
\end{eqnarray}
The trace of the energy-momentum tensor vanishes, meaning that the scalar field is conformally coupled to gravity and the resulting potential preserves the conformal invariance. The second derivative of the gravitational model should be positive in order to be stable and for the choice of the $ f(R)$  function (\Ref{F2}), we have
\begin{equation} f_{RR} = \frac{\alpha }{2 (R-4 \Lambda )^{3/2}}~, \end{equation}
which is always positive for $\alpha >0$.
The horizons can be obtained analytically. For completeness we give the posssibly positive roots of the metric function which for appropriate relations between the constants represent black hole horizons.
\begin{eqnarray}
r_1 &=&\frac{1}{4} \left(\sqrt{\frac{\alpha ^2 \left(\Lambda  \sqrt[3]{L}+2\right)^2-3 \Lambda }{\alpha ^2 \Lambda ^2 \sqrt[3]{L}}}-2 \sqrt{\frac{2}{\Lambda }+\sqrt[3]{L} \left(-\frac{4}{\alpha  \Lambda  \sqrt[3]{L} \sqrt{\frac{\alpha ^2 \left(\Lambda  \sqrt[3]{L}+2\right)^2-3 \Lambda }{\alpha ^2 \Lambda ^2 \sqrt[3]{L}}}}-\frac{1}{4}\right)+\frac{\frac{3 \Lambda }{\alpha ^2}-4}{4 \Lambda ^2 \sqrt[3]{L}}}\right)~,\\
r_2 &=& \frac{1}{4} \left(2 \sqrt{\frac{2}{\Lambda }+\sqrt[3]{L} \left(\frac{4}{\alpha  \Lambda  \sqrt[3]{L} \sqrt{\frac{\alpha ^2 \left(\Lambda  \sqrt[3]{L}+2\right)^2-3 \Lambda }{\alpha ^2 \Lambda ^2 \sqrt[3]{L}}}}-\frac{1}{4}\right)+\frac{\frac{3 \Lambda }{\alpha ^2}-4}{4 \Lambda ^2 \sqrt[3]{L}}}-\sqrt{\frac{\alpha ^2 \left(\Lambda  \sqrt[3]{L}+2\right)^2-3 \Lambda }{\alpha ^2 \Lambda ^2 \sqrt[3]{L}}}\right)~,\\
r_3 &=& \frac{1}{4} \left(\sqrt{\frac{\alpha ^2 \left(\Lambda  \sqrt[3]{L}+2\right)^2-3 \Lambda }{\alpha ^2 \Lambda ^2 \sqrt[3]{L}}}+2 \sqrt{\frac{2}{\Lambda }+\sqrt[3]{L} \left(-\frac{4}{\alpha  \Lambda  \sqrt[3]{L} \sqrt{\frac{\alpha ^2 \left(\Lambda  \sqrt[3]{L}+2\right)^2-3 \Lambda }{\alpha ^2 \Lambda ^2 \sqrt[3]{L}}}}-\frac{1}{4}\right)+\frac{\frac{3 \Lambda }{\alpha ^2}-4}{4 \Lambda ^2 \sqrt[3]{L}}}\right)~,
\end{eqnarray}
where
\begin{equation} L=\sqrt{\frac{88 \alpha ^2 \Lambda -80 \alpha ^4+27 \Lambda ^2}{\alpha ^6 \Lambda ^5}}+\frac{14}{\alpha ^2 \Lambda ^2}-\frac{8}{\Lambda ^3}~.\end{equation}

\begin{figure}[H]
\centering
 \includegraphics[width=.40\textwidth]{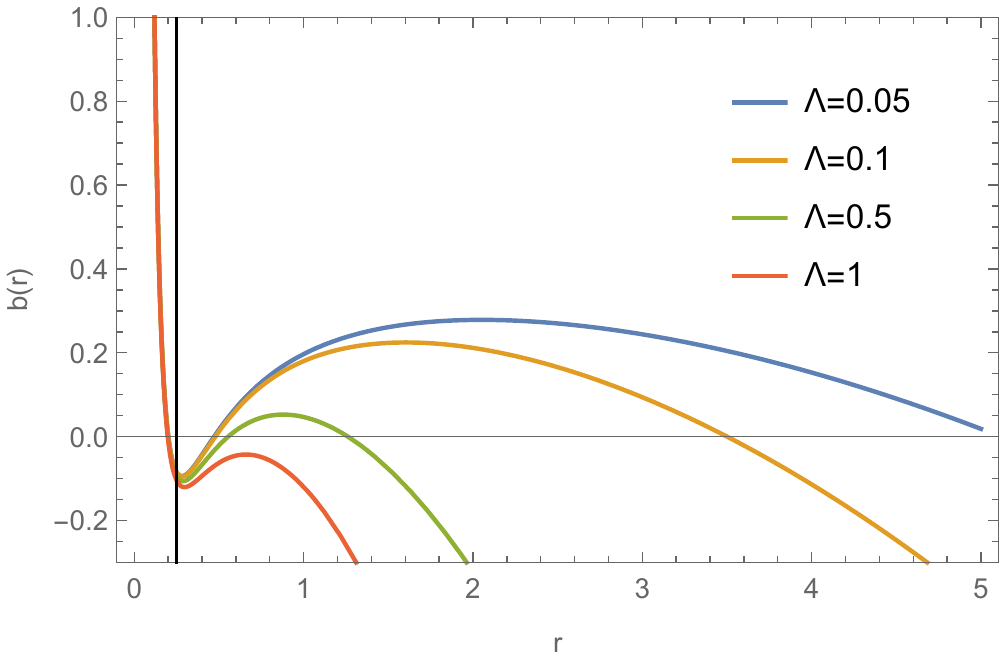}
     \includegraphics[width=.40\textwidth]{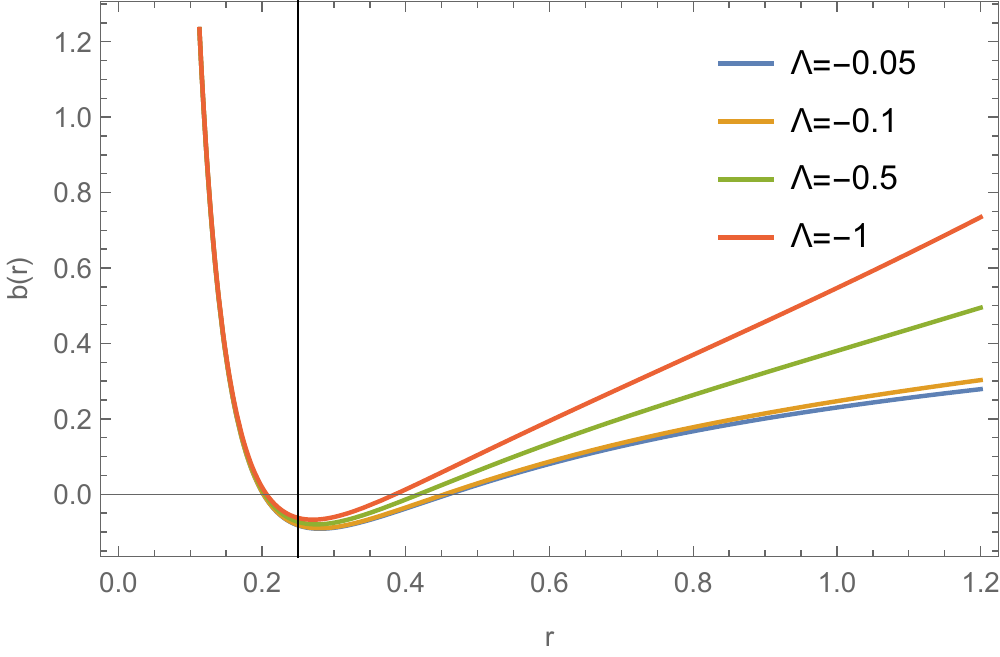}
\caption{The metric function $b(r)$ while changing the cosmological constant, where we've set $\alpha=1$. The black vertical line represents the position of the divergence  of the scalar field. The left plot represents dS spacetimes, while the right AdS spacetimes.}\label{cosmo_metrics} \label{cosmo_metrics_phi}
\end{figure}

In FIG. \Ref{cosmo_metrics_phi} we plot the metric function (\ref{cosmo_b}). For appropriate relations between the parameter $\alpha$ and the cosmological constant,  the positive cosmological constant case gives dS spacetimes with two black hole horizons and one cosmological horizon,  which is what happens in the GR case \cite{Martinez:2002ru}. As in the asymptotically flat case, the divergence of the scalar field is in between of the two black hole horizons.
We compute the Kretschmann scalar for the metric (\Ref{cosmo_b})
\begin{eqnarray}
R^{\alpha\beta\gamma\delta}R_{\alpha\beta\gamma\delta} &=&\frac{8 \Lambda ^2}{3}+\frac{55}{48 \alpha ^2 r^6}-\frac{3}{4 \alpha ^3 r^7}+\frac{63}{512 \alpha ^4 r^8}+\frac{4}{3 \alpha  r^5}+\frac{4 \Lambda }{3 r^2}+\frac{1}{r^4}~,\\
\lim_{r\rightarrow 0} R^{\alpha\beta\gamma\delta}R_{\alpha\beta\gamma\delta} &\rightarrow& \infty~.
\end{eqnarray}
The square of the Weyl tensor remains the same (\Ref{weyl}) as expected, since the cosmological constant does not appear in the Weyl tensor.

To study thermodynamics we consider that the black hole horizon for the metric (\Ref{cosmo_b}) is the largest positive root of the metric function
\begin{equation}r_+ = \text{Root}\left(9-32 \alpha  r \left(2 \alpha  \Lambda  r^3-3 \alpha  r+2\right)\right)~. \end{equation}
Now using the relations (\Ref{T}), (\Ref{S}) we compute the Hawking temperature and the Bekenstein-Hawking entropy as
\begin{eqnarray}
T(r_+) &=&\frac{32 \alpha  r_+ \left(1-2 \alpha  \Lambda  r_+^3\right)-9}{384 \pi  \alpha ^2 r_+^3}~, \label{cosmo_T}\\
S(r_+) &=&\pi  r_+^2 \left(+1-\alpha  r_+-\frac{1}{\left(1-4 \alpha  r_+\right){}^2}\right)~. \label{entropy1}
\end{eqnarray}
In order to make sure that $r_+$ represents the black hole horizon we solve the metric function for the modified gravity parameter $\alpha$ and we obtain
\begin{equation} b(r_+) =0 \to \alpha_{1,2}= \frac{-8 r_+ \pm \sqrt{2} \sqrt{18 \Lambda  r_+^4+5 r_+^2}}{8 \left(2 \Lambda  r_+^4-3 r_+^2\right)}~, \label{alphas}\end{equation}
where the $\alpha_1 $ represents the solution with the plus sign. The modified gravity parameter is positive and therefore both roots of the above equation should be positive. For dS spacetimes $\alpha_{1,2}$ are both positive when the cosmological constant satisfies $ 0<\Lambda <\frac{3}{2 r_+^2}$. For AdS spacetimes, $\alpha_{1,2}$ are both positive when the cosmological constant satisfies $ -\frac{5}{18 r_+^2}\leq \Lambda <0$.
Now, we substitute the expressions for $\alpha$ back to the temperature and we have

\begin{equation} T(r_+)_{1,2} = -\frac{36 \Lambda  r_+^3 \pm 4 \sqrt{2} \sqrt{r_+^2 \left(18 \Lambda  r_+^2+5\right)}+5 r_+}{108 \pi  r_+^2}~. \end{equation}

For AdS spacetimes, the above temperature is positive when the cosmological constant and the horizon satisfy $-\frac{5}{18 r_+^2}\leq \Lambda <-\frac{1}{4 r_+^2}$ which also gives a positive $\alpha$.

For dS spacetimes, the temperature is positive for $\alpha_2$ when the cosmological constant satisfies: $0<\Lambda <\frac{5}{12 r_+^2}$ and is negative for $\alpha_1$.

From equation (\Ref{entropy1}) we can see that regardless of the sign and value of the cosmological constant, the entropy is always negative, for any $r_+>0, \alpha >0$.

\subsubsection{Solution with Charge}

We have seen in the asymptotically flat case that the addition of the Maxwell term in the action, results in positive entropy when the constants of our solution satisfy particular relations. Therefore we consider the action (\Ref{charged_action}) and the gravitational model  (\Ref{F2}) and solving the Einstein-Klein-Gordon-Maxwell system we find that the only functions that change are
\begin{eqnarray}
V(r) &=& \frac{\left(4 \alpha ^2+\Lambda \right) \left(32 \alpha ^2 Q^2-3\right)}{3 (1-4 \alpha  r)^4}~,\\
V(\phi) &=& \frac{\phi ^4 \left(4 \alpha ^2+\Lambda \right)}{384 \alpha ^2 Q^2-36}~,
\end{eqnarray}
while the scalar field is now given by (\Ref{charged_phi}) and the metric function, the Ricci scalar and the functions $f_R$ and $f(r)$ are given by (\Ref{cosmo_b}), (\Ref{cosmo_R(r)}), (\Ref{cosmo_f(r)}) and   (\Ref{cosmo_f_R}) respectively.
The temperature of the solution is given by (\Ref{cosmo_T}) and we discussed in detail the possibillity of positive temperature, while the entropy is given by
\begin{equation} S(r_+) = \pi  r_+^2 \left(1 -\alpha  r_+ + \frac{32 \alpha ^2 Q^2-3}{3 (1-4 \alpha  r_+)^2}\right)~. \label{S_Q_L}\end{equation}
We will now follow the same procedure as in the uncharded case. In order to ensure that the entropy (\Ref{S_Q_L}) is indeed the entropy at the black hole horizon we solve the metric function for the parameter $\alpha$ and substitute back to the entropy.
We find that for particular range of values for the electric charge and the cosmological constant, the entropy is positive for positive $\alpha$. The inequalities are complicated and therefore, we will only give the simplest one in order to illustrate the fact that we can have positive entropy. For $\alpha_2$ and $-\frac{5}{18 r_+^2}\leq \Lambda <0$ (AdS spacetime) the entropy is positive when the electric charge and the black hole horizon are related through
\begin{equation} 4 Q+\sqrt{3} \sqrt{\frac{r_+^2 \left(5 \left(\sqrt{36 \Lambda  r_+^2+10}-4\right)-4 \Lambda  r_+^2 \left(\sqrt{36 \Lambda  r_+^2+10}-2\right)\right)}{2 \Lambda  r_+^2-3}}<0~, \end{equation}
or,
 \begin{equation}
4 Q>\sqrt{3} \sqrt{\frac{r_+^2 \left(5 \left(\sqrt{36 \Lambda  r_+^2+10}-4\right)-4 \Lambda  r_+^2 \left(\sqrt{36 \Lambda  r_+^2+10}-2\right)\right)}{2 \Lambda  r_+^2-3}}~.
\end{equation}
For these particular values we also have positive temperature.

\section{Conclusions}
\label{sect4}

In this work we studied  $f(R)$ gravity theories in the presence of matter. We considered a scalar field non-minimally coupled to gravity in the context of $f(R)$ theories. We derived asymptotically flat or (A)dS exact black hole solutions with dynamic Ricci curvature. We also studied the thermodynamics of these black hole solutions. We calculated the temperature and the entropy and because of the presence of the non-minimal coupling between the scalar field and the scalar curvature the area law of the Bekenstein-Hawking entropy is modified resulting to some interesting properties for the entropy.

We first considered the case where $f(R) = R -2\alpha \sqrt{R}$. The parameter $\alpha$ which has the dimension of inverse length, introduces a non-linear correction term to the Ricci scalar $R$. This parameter induces a charge in the scalar field function which however is not independent of the black hole mass and therefore cannot be detected asymptotically, failing in this way to dress the black hole with a primary hair.  Calculating the temperature of the black hole solution we find that it is positive for any value of the parameter $\alpha$. However, the entropy, having a contribution from the scalar field, is negative regardless of the value of the parameter $\alpha $. To cure this problem we introduced an electromagnetic field. Then we found that because of the conformal invariance, the electric charge does not appear in the metric function giving charge to the black hole solution, it appears however in the scalar field function making the entropy positive.

We then introduced a cosmological constant in the $f(R)$ function $f(R) =R -2 \Lambda -2 \alpha  \sqrt{R-4 \Lambda }$. We found dS and AdS black hole solutions depending on the sign of the cosmological constant, while the scalar field has the same behavior with the asymptotically flat case, it cannot give hair to the black hole solutions. If we introduce electric charge this charge appears in the scalar function and the interplay between the cosmological constant and the electric charge  can result to a positive entropy.

It is known that in order to have a black hole with a non-minimally coupled scalar field regular at the horizon, a scale has to be introduced in the theory \cite{Martinez:2002ru,Khodadi:2020jij, Priyadarshinee:2021rch}, which, so far, has been considered to be a cosmological constant. In this work we showed that this not the only possibility. A scale in the form of a non-linear correction term in the gravitational part of the action, also results to a regular scalar field at the black hole event horizon. This would be interesting to be further investigated. One could also consider non-linear electrodynamics, instead of Maxwell electrodynamics to see if it is possible to construct regular black hole solutions (without a singularity). In our case, in order to respect the conformal invariance, the $f(R)$ model and the Ricci scalar are fixed in a particular way and regular black hole solutions cannot be found since Ricci scalar is divergent at the origin and so are the other curvature scalars. The introduction of non-linear electrodynamics will break the conformal invariance making it hopeful to construct regular black holes as it was done in  \cite{Rodrigues:2015ayd, Rodrigues:2016fym}.

 It would be interesting to extent this work to the study of other compact objects in the $f(R)$ gravity theories. The conformally coupled scalar field can provide the matter content of these theories. Then the astrophysical observations in this strong gravity regime may give important information on the departure  from GR restricting the parameter $\alpha$ which expresses the deviation from the Ricci scalar $R$.

\end{document}